\documentclass[final,authoryear,5p, twocolumn]{elsarticle}

\usepackage{graphicx}
\usepackage{amssymb}

\usepackage[ps2pdf,%
colorlinks=true,%
pdfauthor={M\'etrailler et al.},%
pdftitle={Data-Driven Modelling of the Van Allen Belts}%
]{hyperref}
\usepackage{breakurl}
\usepackage[latin1]{inputenc}
\usepackage[T1]{fontenc}
\usepackage{wasysym}        % To enable the use of the diameter symbole
\usepackage[labelformat=simple]{subcaption}     % To enable the subfigures

\usepackage{dblfloatfix}    % To enable figures at the bottom of page
\usepackage{graphbox}       % To enable the use of align command in includegraphics
\usepackage[capitalise,noabbrev]{cleveref}
\usepackage{xspace}
\usepackage{soul}

\graphicspath{{figures/}}

\newcommand{\xmm}{{\sl XMM-Newton}\xspace}
\newcommand{\integ}{{\sl INTEGRAL}\xspace}
\newcommand{\rearth}{\,$R_{\rm{E}}$\xspace}

\journal{Advances in Space Research}

\begin{document}

%% Frontmatter
\begin{frontmatter}


% Use the tnoteref command within \title and fnref within \author or \address for footnotes;
% use the corref command within \author for corresponding author footnotes;
% \author{Name\corref{cor1}\fnref{label2}}
% \address{Address\fnref{label3}}

\title{Data-Driven Modelling of the Van Allen Belts: The 5DRBM Model for Trapped Electrons}

% Use optional labels to link authors explicitly to addresses:

\fntext[footnote1]{Swiss National Trainee Program, Swiss Space Center, \'Ecole Polytechnique F\'ed\'erale de Lausanne (EPFL), PPH 338, Station 13, CH-1015 Lausanne, Vaud, Switzerland}
\fntext[footnote2]{European Space Agency - ESA/ESAC, Camino Bajo del Castillo, 28692 Villanueva de la Ca\~nada, Madrid, Spain}
\fntext[footnote3]{European Space Agency - ESA/ESTEC, Keplerlaan 1, 2201 AZ Noordwijk, Netherlands}
\fntext[footnote4]{European Space Agency - ESA/ESOC, Robert-Bosch-Strasse 5, D-64293 Darmstadt, Germany}

\author{Lionel M\'etrailler\corref{cor}\fnref{footnote1}}
%\address{European Space Astronomy Centre - ESA/ESAC, Villanueva de la Ca\~nada, Madrid, Spain}
\cortext[cor]{Corresponding author}
\ead{lionel.metrailler@esa.int}

% Url can be given like this:
% \ead[url]{http://www.elsevier.com/wps/find/authorsview.authors/latex}

\author{Guillaume B\'elanger\fnref{footnote2}}
%\address{European Space Astronomy Centre - ESA/ESAC, Villanueva de la Ca\~nada, Madrid, Spain}
\ead{guillaume.belanger@esa.int}
 
\author{Peter Kretschmar\fnref{footnote2}}
\ead{peter.kretschmar@esa.int}

\author{Erik Kuulkers\fnref{footnote3}}
\ead{erik.kuulkers@esa.int}

\author{Ricardo Pérez Martínez\fnref{footnote2}}
\ead{Ricardo.Perez.Martinez@esa.int}

\author{Jan-Uwe Ness\fnref{footnote2}}
\ead{jan.uwe.ness@esa.int}

\author{Pedro Rodriguez\fnref{footnote2}}
\ead{pedro.rodriguez@esa.int}

\author{Mauro Casale\fnref{footnote2}}
\ead{mauro.casale@esa.int}

\author{Jorge Fauste\fnref{footnote2}}
\ead{jorge.fauste@esa.int}

\author{Timothy Finn\fnref{footnote4}}
\ead{timothy.finn@esa.int}

\author{Celia Sanchez\fnref{footnote2}}
\ead{celia.sanchez@esa.int}

\author{Thomas Godard\fnref{footnote4}}
\ead{thomas.godard@esa.int}

\author{Richard Southworth\fnref{footnote4}}
\ead{richard.southworth@esa.int}

\address{European Space Astronomy Centre - ESA/ESAC, Villanueva de la Ca\~nada, Madrid, Spain}

\begin{abstract}
The magnetosphere sustained by the rotation of the Earth's liquid iron core traps charged particles, mostly electrons and protons, into structures referred to as the Van Allen belts.
These radiation belts, in which the density of charged energetic particles can be very destructive for sensitive instrumentation, have to be crossed on every orbit of satellites traveling in elliptical orbits around the Earth, as is the case for ESA's \integ and \xmm missions.
This paper presents the first working version of the 5DRBM-e model, a global, data-driven model of the radiation belts for trapped electrons.
The model is based on in-situ measurements of electrons by the radiation monitors on board the \integ and \xmm satellites along their long elliptical orbits for respectively 16 and 19 years of operations.
This model, in its present form, features the integral flux for trapped electrons within energies ranging from 0.7 to 1.75 MeV.
Cross-validation of the 5DRBM-e with the well-known AE8min/max and AE9mean models for a low eccentricity GPS orbit shows excellent agreement, and demonstrates that the new model can be used to provide reliable predictions along widely different orbits around Earth for the purpose of designing, planning, and operating satellites with more accurate instrument safety margins.
Future work will include extending the model based on electrons of different energies and proton radiation measurement data.
\end{abstract}

\begin{keyword}
%first keyword \sep second keyword \sep more keywords
% The keywords are taken from a fixed list of keywords supplied by the Journal
Van Allen belts \sep Radiation belt modelling \sep Trapped particles \sep Radiation Environment \sep Space environment \sep Space weather
\end{keyword}

\end{frontmatter}

\parindent=0.5 cm

%%%%%%%%%%%%%%%%%%%%%%%%%%%%%%%%%%%%%%%%%%%%%%%%%%%%%%%%%%%%%%%%%%%%%%%%%%%%%
%% Main text

\section{Introduction}

Only theorised before space exploration began \citep{Stormer37}, the Earth Radiation Belts were discovered for the first time in 1958 with the very first US satellite, Explorer 1.
These first in-situ measurements allowed Mr.\ James Van Allen, payload specialist of the Explorer missions, to discover successively: the inner proton belt (Explorer 1 \& 3), the inner electron belt (Explorer 4) and the outer electron belt (Pioneer 3) \citep{VanAllen59}.
Consequently, these high radiation regions surrounding the Earth are known as the Van Allen Belts (VAB).

These radiation belts are the result of the complex interaction between the Earth's magnetosphere and the interplanetary medium mainly driven by the solar wind that moves energetic charge-carrying electrons and protons with it.
These particles are trapped in the magnetosphere forming high radiation torus-shaped regions around the Earth's magnetic axis.
The Van Allen Belts extend from altitudes from $\sim$1,000km (0.2\rearth) to more than $\sim$60,000km (10\rearth).

\begin{figure*}[!b]
    \centering
    \begin{subfigure}[t]{0.48\textwidth}
        \centering
        \includegraphics[width=0.91\textwidth]{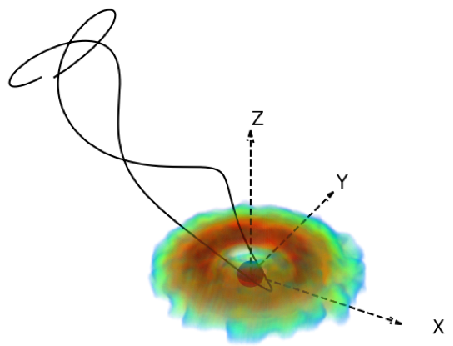}
        \caption*{\integ}
        %\label{fig:intorb}
    \end{subfigure}%
    ~
    \begin{subfigure}[t]{0.48\textwidth}
        \centering
        \includegraphics[width=0.65\textwidth]{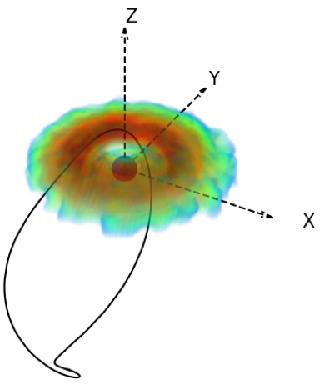}
        \caption*{\xmm}
    \end{subfigure}
    \caption{\xmm and \integ complementary orbits in the Solar Magnetic (SM) frame. The orbits are not elliptical due to the daily oscillation and yearly rotation of the SM frame with respect to the Geocentric Equatorial Inertial (GEI) frame. The Earth is plotted to scale as a sphere in the centre, and the VAB are shown as the colour-shaded regions giving a qualitative idea of the radiation intensity: the blue corresponds to a low radiation density region and the red is a high density region. Positive $x$ always points towards the Sun.\label{fig:intorb-xmmorb}}
\end{figure*}

Modelling these highly dynamic radiation belts is important for space-borne activities, both manned and unmanned, in the near-Earth environment.
Theoretical models of the belts can be built on simplified physical equations which describe the movement and behaviour of trapped particles.
The trapped particles dynamics have already been well explained in the seventies \citep{Roederer70} as well as in more recent studies using the Van Allen Probes showing the presence of a temporary third trapped radiation belt \citep{Boyd18}.
The general shape of the VAB is relatively stable except for the occasional appearance of the third temporary belt following strong events in the geomagnetic sphere.
However, in practice, the outer-boundary of the outer electron radiation belt is not constant in time (see \cref{fig:IQRs}).
The VAB can grow rapidly when solar eruptions reach the Earth magnetosphere and then deflate through various trapped particle losses acting at different temporal and spatial scales \citep{Baker18,Vassiliadis04,Thorne10}.
These factors make it difficult to simulate the outer part of the trapped electron belt.

In this paper, we use radiation flux measurements recorded by radiation monitors on board two spacecraft, \integ and \xmm.
Both have been on highly elliptical orbits around the Earth for respectively 16 and 19 years probing the Southern and Northern hemispheres, respectively.
Using these measurements, we construct a dynamical 3D volume model for the VAB based on the electron radiation detected in the 0.7--1.75 MeV energy range.

\begin{table*}[!b]
    \centering
    \caption{Main characteristics of the on-board Si scintillators radiation monitors. The electron energy range corresponds to the one available on the ESA Open Data Interface (ODI) server.}
    \vspace{0.1cm}
    \begin{tabular}{l||c|c}
        \hline
        \textbf{Radiation monitors}&\textbf{\integ}&\textbf{\xmm}\\
        \hline
        \hline
        \textit{Side shield} & 4.2mm Ta \& 5mm Al & 5mm Al \\
        \textit{Front shield} & 0.65mm Al & 0.02mm Be\\
        \textit{FOV/Opening} & $\pm20^\circ$ conical & 3mm \diameter, 1sr, $\pm32.8^\circ$ \\
        \textit{Thickness} & 0.5mm Si & 0.5mm Si\\
        \textit{Active Area} & 25mm$^2$ & 85mm$^2$\\
        \textit{e$^-$ Energy Range} & 0.65MeV - 2.18MeV & 0.13MeV - 1.87MeV\\
    \hline
    \end{tabular}
    \label{table:radmon}
\end{table*}

There are currently a few global dynamical models for the VAB but they do not easily meet the needs that long term planning requires for scientific missions such as \integ or \xmm.
For example, the British Atlantic Survey - Radiation Belt model (BAS-RBM) \citep{Glauert14} is a global dynamic model that simulates the high energy electron population (>500 keV) of the radiation belts taking into account effects such as the changing solar activity and wave-particle interactions.
It is mainly based on the Fokker-Planck equation and satellite data.
This model is very good for simulations of past VAB states and detailed forecasts on time scales of hours to days.

Another example of a global model is the Global Radiation Earth ENvironment (GREEN) model \citep{Sicard18}.
GREEN is a global model using various global and local models to obtain the most reliable value at each point in space.
It is very good for detailed local simulations. For a global 3D view of the VAB, GREEN might not be the easiest model to use for long term forecasts either.

NASA's AE8min/max and AP8min/max models \citep{SawyerAP876,VetteAE891} for electrons and protons, respectively, are the most well-known and have been used since the 1970s.
More recently, the new IRENE-AE9/AP9 models have been issued \citep{OBrien17,Johnston15}.
In these models, the trapped charged particle populations are treated in McIlwain's ($B, L$) or ($\alpha_{eq}, L^{\star}$) coordinate system \citep{McIlwain61}.
This system tries to use the symmetries inherent to trapped particles behaviour in a magnetosphere in order to increase the measurement sampling and, consequently, have a better statistical validity.
This results in a 2D view of the Radiation Belts, where only a section of the radiation torus is visible.
The Electron Slot Region Radiation Environment Model \citep{Sandberg14} is the closest to the model presented in this work, however it also uses the ($\alpha_{eq}, L^{\star}$) coordinate system.

A data-driven model based purely on measurements in the Cartesian 3D space around the Earth will naturally include these asymmetries.
The best reference frame for such a model is one in which the global structure of the belts is mostly static.
This is the case in the Solar Magnetic (SM) reference frame where the $z$-axis is parallel to the Earth's magnetic dipole (11$^{\circ}$ degrees tilt with respect to its rotation axis) with its positive direction towards the northern hemisphere (south magnetic pole);
the $x$-axis is defined in the plane given by the $z$-axis and the Earth-Sun line with the positive direction towards the Sun;
and the $y$-axis is defined to have an orthogonal system.
Trajectories plotted in this frame will not be elliptical due to the daily oscillation and yearly rotation of the SM frame with respect to the Geocentric Equatorial Inertial (GEI) frame.

The model presented in this paper uses a new and simpler approach to visualize the VAB by using the SM 3D reference frame to process the data, create, and use the model.
It implies a 3D view of all possible asymmetries detected in the belts. As seen in the $xz$-section view of the static VAB model presented here (see \cref{fig:step4}), the belt appears to have a stronger radiative core on the night side (negative $x$-direction) and a slightly broader section on the day side (positive $x$-direction).
This justifies the use of the 3D SM system instead of McIlwain's.
The caveat of using the full 3D space in the SM reference frame resides in the sampling of the 3D space which in that case is naturally lower than with McIlwain's reference system.
The accumulation of 18 years of data in addition to the complementarity of \integ's and \xmm's orbital configurations (see \cref{fig:intorb-xmmorb}) allow for a global coverage of the main parts of the VAB as explained in \cref{sec:volume-model}.

This paper presents a new, empirical data-driven model of the Earth's radiation belts denoted \emph{5DRBM-e}, where \emph{5D} represents the model's five dimensions (three for the spatial position, one for time, and one for the intensity of radiation);
\emph{RBM} stands for \emph{Radiation Belts' Model}; and \emph{-e} stands for \emph{electrons}.
\cref{sec:radiation-monitors} presents the basic characteristics of the radiation monitors from which the data are taken, \cref{sec:volume-model} describes how the volume model is built, and \cref{sec:model-validation} shows how it can be applied in practice, and how it compares to the AE8min/max models.

\section{The radiation monitors}
\label{sec:radiation-monitors}

There are currently two active ESA missions that cross the VAB sampling almost all their structures. They are \xmm and \integ, which were launched respectively in 1999 and 2002.
These crossings along the 2-3 days orbits of the spacecraft allow for a scanning of the belts from their outer boundaries down to approximately 2,500 km above the Earth's surface with \integ data.
This, however, limits reliability at altitudes below 3,000 km, something that could be addressed in the future by including LEO measurements in the construction of the model.

The \integ Radiation Environment Monitor, IREM \citep{Hajdas03} and the \xmm European Photon Imaging Camera Radiation Monitor, EPIC RM or ERM \citep{Boer95} gather in-situ radiation data continuously along the spacecraft's orbit.
Remarkably, not only have these two spacecraft gathered more than 16 years of contemporaneous radiation measurements, but as is shown in \cref{fig:intorb-xmmorb}, their orbits scan different parts of the VAB and are thus complementary in increasing the coverage of the belts.

\cref{table:radmon} shows the main characteristics of the two radiation monitors, IREM and ERM, which are in fact significantly different instruments.
One needs to be careful when combining the data of both missions. Cross-calibration is necessary.
The calibrated radiation flux measurements are available through the ESA Open Data Interface (ODI) server\footnote{The ESA Open Data Interface can be accessed directly at \\ \url{https://spitfire.estec.esa.int/trac/ODI/wiki/ODIv5}} using a simple Python client. The ESA ODI provides ready-to-use space environment data from several different missions, including differential omnidirectional and distinct electron and proton fluxes in different energy ranges for IREM \citep{Mohammadzadeh03,Sandberg12} and ERM.

The on-board radiation monitors are intended to trigger the shutdown and thus protection of the scientific payload instruments in case of excessive radiation \citep{Gonzalez18}.
In every revolution, an instrument window is defined during which the radiation level is expected to be low enough to use the instruments.
The endpoints of this instrument window are predicted based on a simple model of the radiation environment.
The safety of scientific payload relies on ensuring instruments are operating in safe, low-radiation conditions.
The monitors ensure that during the instrument window, the radiation remains below the operational threshold.
Modelling radiation accurately is important for the safety and lifetime of these instruments, but it also allows for smoother science operations by relieving the burden of having to perform re-activation sequences for instruments following an emergency shutdown caused by unpredicted high radiation.

For \xmm, the prediction is based on a 3D surface model of the outer boundary of the electron belt developed in 2004 by Mauro Casale and Jorge Fauste of the \xmm science operations centre \citep{Casale04}.
This model uses the first years of the mission to estimate the outer boundary of the main electron radiation belt.
It works well and is still used today for the science payload mission planning.
But because it was tailor-made for \xmm, this model cannot be applied to another mission.
In addition, because it is a surface model that gives estimates of the boundary of the outer shell of the electron belt, it does not provide radiation profiles along the length of the crossings through the belts.

For \integ, the safe belt entry and exit altitudes are evaluated every month based on the measurements of the two previous years.
A simple sinusoidal fit is applied to the measured altitude at a fixed radiation level.
Based on this fit and accounting for some margins, the available window for observations results from these defined entry and exit altitudes at which the instruments are switched off and on.
Given the orbit's stability, this solution works very well in terms of instrument safety and ease of operations. However, due to smoothing of shorter term dynamic radiation behaviour, scientific observation time is lost.
Moreover, the predictions can be unreliable when the spacecraft crosses different parts of the belts, something that became more common following the orbit changes in 2015 (see \cref{subsec:self-consistency-check}).
A global 3D volume model could predict the different altitudes for successive revolutions and thus yield more reliable estimates.
The limitations of the current methods used to ensure safe operations on \xmm and \integ were the main drivers for the work presented here.

\begin{figure*}[!t]
	\includegraphics[width=1\textwidth]{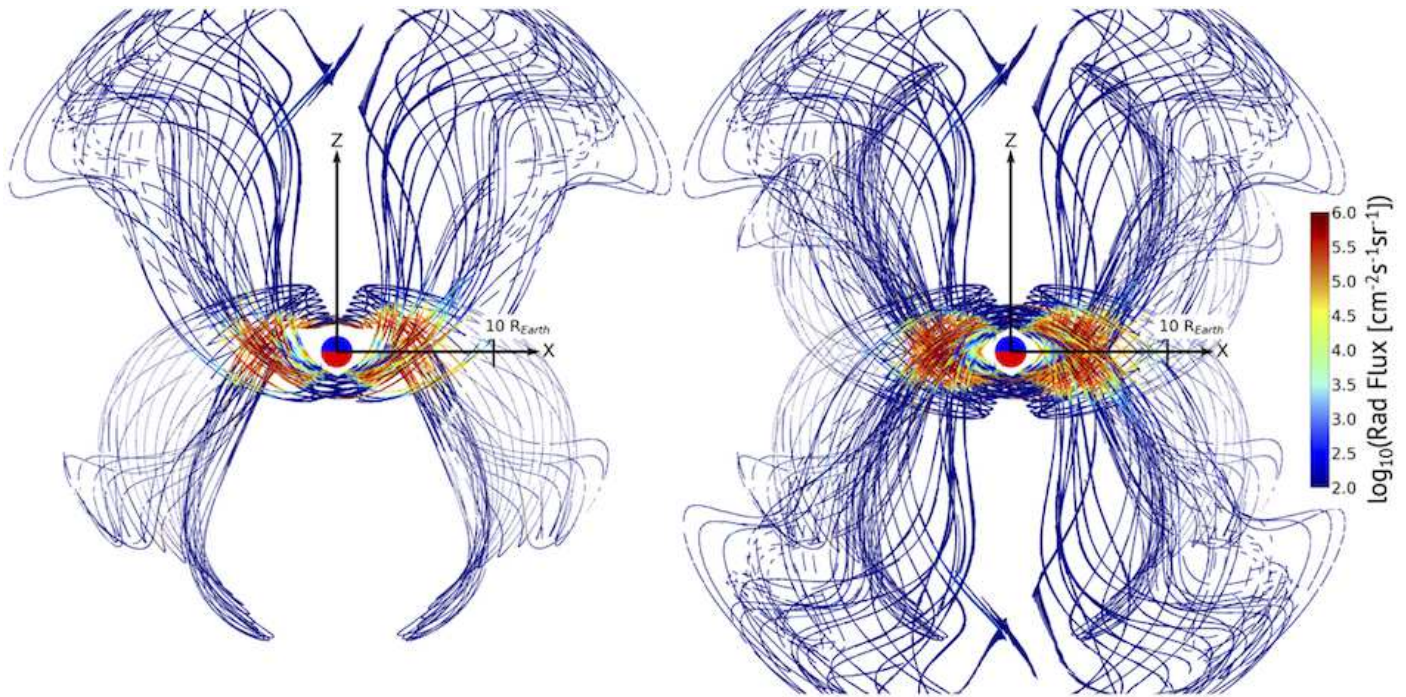}
	\caption{The left plot shows the real tracks in the $x-z$ plane of the SM frame, \integ (upwards) and \xmm (downwards), colour-coded with the cross-calibrated flux. The right figure pictures the duplicated tracks after the mirroring step.\label{fig:tracks}}
\end{figure*}

\section{Building a time-dependent volume model of the Van Allen belts}
\label{sec:volume-model}

A reliable simulation of the radiation environment around the Earth requires a global understanding of the radiation belts' dynamics and the amount of data accumulated using \integ and \xmm can help us to have a phenomenological knowledge on this.
The 5DRBM-e model is built in a modular way from a static model over which time-dependent functions can be applied to account for the dynamics.
Time-dependent deviations from the static model are quantified using the InterQuartile Range (IQR) of radiation flux measurements in cells over the volume of the belts, as defined in the model.
The IQRs can help to understand where the VAB are strongly variable, and computing them for different time scales can give an idea of the time evolution of the belts' general shape.

All the differential omnidirectional electron fluxes with their corresponding positions and time-stamps are taken from the ODI server: these are measured for \integ at [0.7, 0.78, 1.125, 1.27, 1.435, 1.615, 1.75] MeV and for \xmm at [0.7, 0.825, 0.985, 1.2, 1.41, 1.58, 1.75] MeV. Hence, an integration from 0.7 to 1.75 MeV is performed to have one homogeneous data point per position and time-stamp.\footnote{The energy binning for each mission is different, and thus an interpolation in the electron energy spectrum is done to have the same bins edges resulting in an overlapping range from 0.7 to 1.75 MeV instead of 0.65 to 1.87 MeV one can see in \cref{table:radmon}.}

The data are then cleaned through simple sigma clipping and smoothing to produce a data set with no major outliers, which can result from faulty measurements, but also from a wrong calibration process due to, for example, changing in sampling rates (1--2 wrong successive data points, 2 times per orbit for \xmm).
Smoothing the data helps to remove the very short variations (of the order of tens of minutes) or data spikes that we don't want to take into account in this model.

Positions are converted to the SM reference frame.
The data sampling rate is standardised to one measurement per minute (averaging the data points every minutes if the sampling rate is higher) in order to have the same spatial density of points for both satellites along their trajectories in the radiation belts.
Because the radiation monitors on each spacecraft are different, radiation measurements have to be cross-calibrated.
Even if the measurements are in some manner calibrated on the ODI server, this cross-calibration is necessary.

The cross-calibration is performed separately for the highest and lowest radiation levels that correspond to the inside and outside of the belts, respectively.
At the low end, the cross-calibration is done by comparing \xmm and \integ radiation measurements made at the same time far away from the VAB.
This implies assuming that outside of the VAB the electron flux is isotropic, which is not strictly accurate, but since the goal is to focus on the belts themselves, this assumption is good enough because the radiation levels outside the VAB are orders of magnitudes lower than inside.
At the high end, the cross-calibration is performed by first creating independent \xmm and \integ models of the VAB, and then comparing the radiation at corresponding points in the two mission-specific volume models.
In this cross-calibration process, IREM is taken as the baseline because the data show a much smoother behaviour and less scattering than for ERM.
These two cross-calibrations at the low and high ends result in a conversion function to be applied to ERM measurements.
To make the transition between the two regimes smooth, the transition regions are smoothed using local spatial averaging.

\subsection{The static volume model}

After cross-calibration, all data points are plotted in the SM frame.
The 3D grid in units of Earth radii, \rearth, is defined on a regular mesh of 0.1\rearth.
The volume of the grid is centered on the Earth, 14\rearth\ in the $z$ direction, and 32\rearth\ in both the $x$ and $y$ directions in order to encompass the whole Van Allen Belts.
Knowing the precise position of each measurement within the 3D volume, the value in each node of the grid can be computed using the nearest data points.
In order to have a full coverage of this volume, all tracks are mirrored with respect to the magnetic equatorial plane, as shown in \cref{fig:tracks}.
This assumes an up/down symmetry in the SM reference frame, which is commonly assumed in magnetospheric studies. Time dependencies can be added later to take into account up/down asymmetries.

\begin{figure*}[!b]
    \centering
    \begin{minipage}{0.79\textwidth}
        \begin{subfigure}[t]{1\textwidth}
            \centering
            \includegraphics[width=0.78\textwidth]{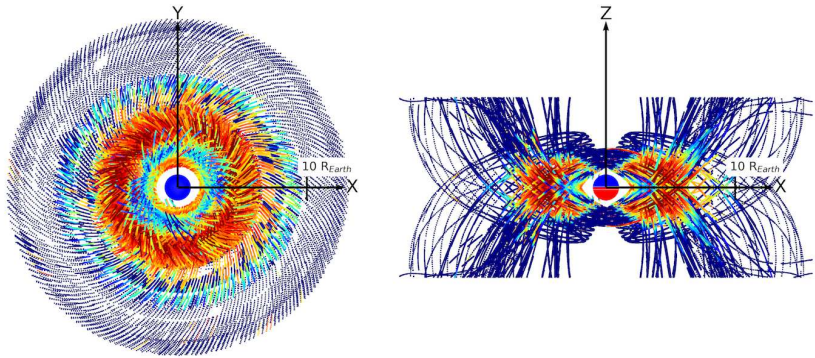}
            \caption{Real tracks in the SM frame colour-coded with the radiation flux\label{fig:step1}}
        \end{subfigure}
        \begin{subfigure}[t]{1\textwidth}
            \centering
            \includegraphics[width=0.78\textwidth]{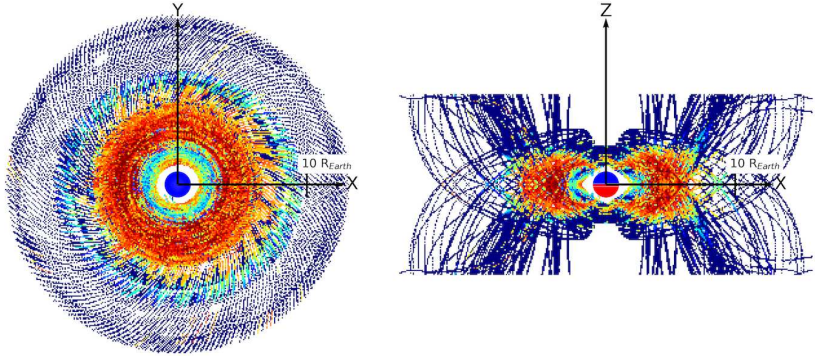}
            \caption{Radiation fluxes assigned to nodes of the 3D grid\label{fig:step2}}
        \end{subfigure}
        \begin{subfigure}[t]{1\textwidth}
            \centering
            \includegraphics[width=0.78\textwidth]{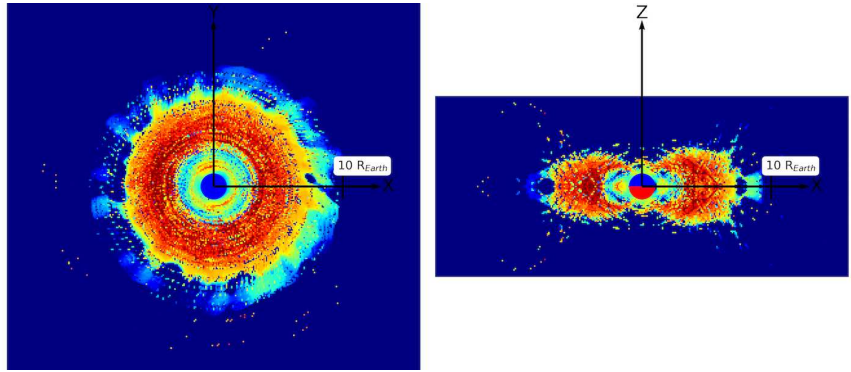}
            \caption{Filled 3D grid using neighbouring nodes\label{fig:step3}}
        \end{subfigure}
        \begin{subfigure}[t]{1\textwidth}
            \centering
           \includegraphics[width=0.78\textwidth]{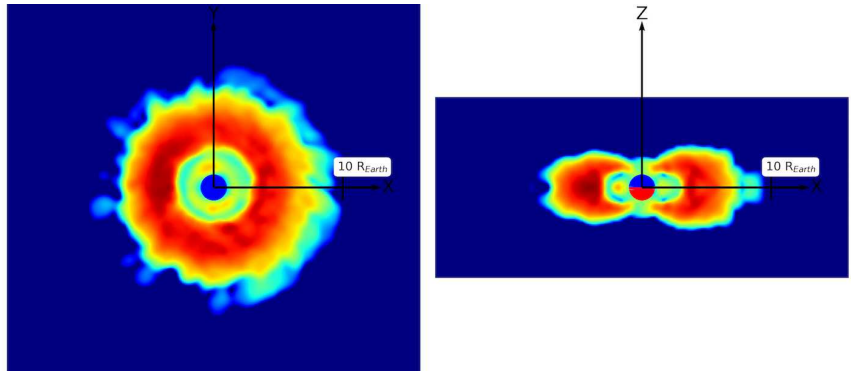}
            \caption{Smoothed 3D grid resulting in the finals 3D static model\label{fig:step4}}
        \end{subfigure}
    \end{minipage}
    \begin{minipage}{0.07\textwidth}
        \centering
        $\vcenter{\hbox{\includegraphics[width=1\textwidth]{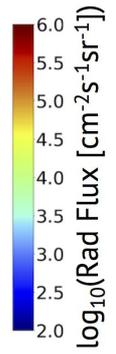}}}$
    \end{minipage}
    \caption{Main building steps of the 3D static model using real data points. The left side of each figure is the $x-y$ plane top view and the right side is the $x-z$ plane side view. Everything is colour-coded with the cross-calibrated electron flux. The 3D grid has a size of 32x32x14 Earth radii. \label{fig:allsteps}}
\end{figure*}

\begin{figure*}[!t]
    \centering
    \begin{minipage}{0.78\textwidth}
        \centering
        \includegraphics[width=0.8\textwidth]{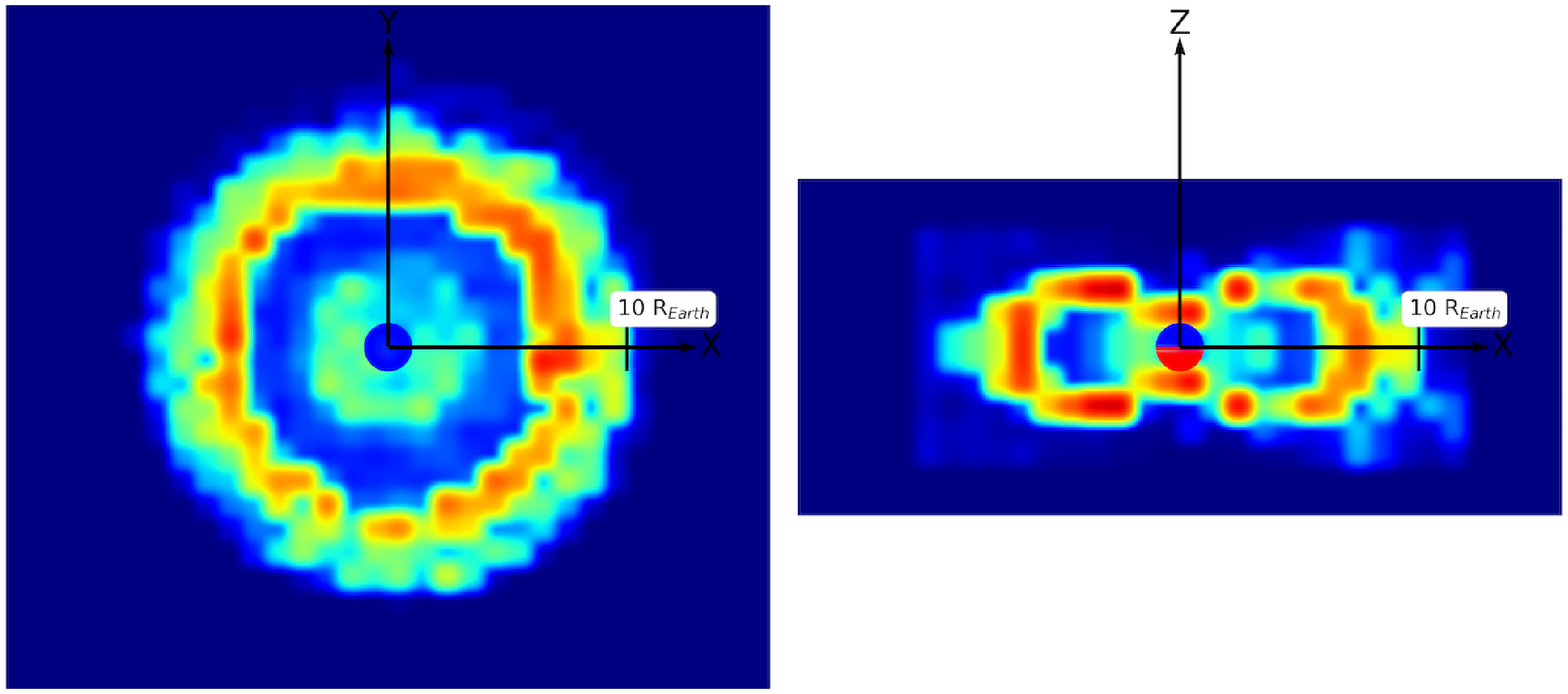}
    \end{minipage}
    \begin{minipage}{0.07\textwidth}
        \centering
       \includegraphics[width=1\textwidth]{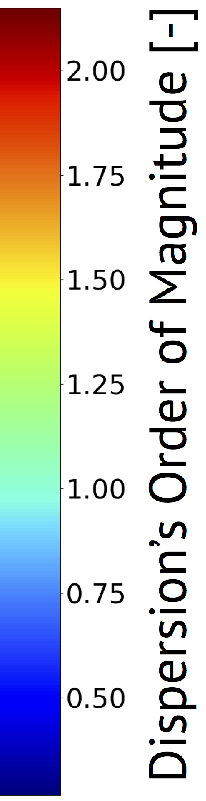}
    \end{minipage}
    \caption{Interquartile ranges 3D map projections. The left side is the top view of the $x-y$ plane, and the right side is the  side view of the $x-z$ plane. High dispersion regions are located mainly at the boundary of the outer belt. The 3D grid has a size of 32x32x14 Earth radii.\label{fig:IQRs}}
\end{figure*}

Mirroring of the tracks increases the number of nodes in which an average radiation value can be computed.
This does not, however, cover the entire volume.
Looking at a 3D rectangle of size $16\times16\times6$\rearth$^3$ centered on the Earth, covering the main VAB regions below 45,000 km altitude, 53\% of all cells of size 0.1\rearth contain at least 1 data point, and half contain 3 or more data points.
This is not enough to build an accurate model down to a resolution of 0.1\rearth.
Consequently, the remaining voids (nodes without an assigned value) in the grid need to be filled by iteratively using the nearest neighbouring nodes.
Four iterations are performed to fill the full 3D grid.
An empty node value is found computing the median of all the neighbouring non-zero nodes in a 0.8\rearth\ box size.\footnote{The box size was optimized as a function of the required resolution and the processing time.}
In the 3D rectangle of size $16\times16\times6$\rearth$^3$ centered on the Earth, more than 95\% of all cells of size 0.8\rearth contain at least 1 data point with a median population per cell of more than 500, which is enough for a good statistical sampling.

\cref{fig:allsteps} illustrates the steps used in the construction of the static model.
Values in the resulting 3D matrix are smoothed to obtain the final static volume model of the trapped electron belts.
A short 3D animation, showing the creation of the static model, has been selected for the ESA INTEGRAL Picture Of the Month (POM) in July 2018\footnote{INTEGRAL POM July 2018: \url{https://www.cosmos.esa.int/web/integral/pom-archive}} and at the same time for the NASA High Energy Astrophysics Picture Of the Week (HEAPOW)\footnote{HEAPOW 2nd of July 2018: \url{https://heasarc.gsfc.nasa.gov/docs/objects/heapow/archive/solar_system/vab_integral.html}}.

It is this static model that constitutes the basis for the more realistic time-dependent model for which the most important element is the variation due to the solar cycle.

\subsection{The time dependence}

The solar cycle on timescales of a few years is very well described by a simple sinusoid with an 11-year period.
A simple sinusoidal fit to the logarithm of the radiation flux measured outside the VAB during the past 18 years, combining \xmm and \integ data, describes the long term variation of the background radiation.
The parameters of this fit are determined during the elaboration of the model, and can then be used to vary the mean background radiation of the static model as a function of time when the model is used for simulations or predictions.\footnote{No external parameters are used to include this solar cycle in the model. Future work could include a more elaborate fitting of the solar cycle, or the use of external solar cycle variables.}
The background radiation is higher during lower solar activity and lower during higher solar activity periods.
This is because this background radiation is mainly composed of galactic cosmic-rays whose density is strongly modulated by the solar cycle in an inversely proportional manner.
The influence of the solar cycle on the VAB is much more complex and has not, at this stage, been implemented.

\begin{figure*}[!t]
	\centering
    \includegraphics[width=0.87\textwidth]{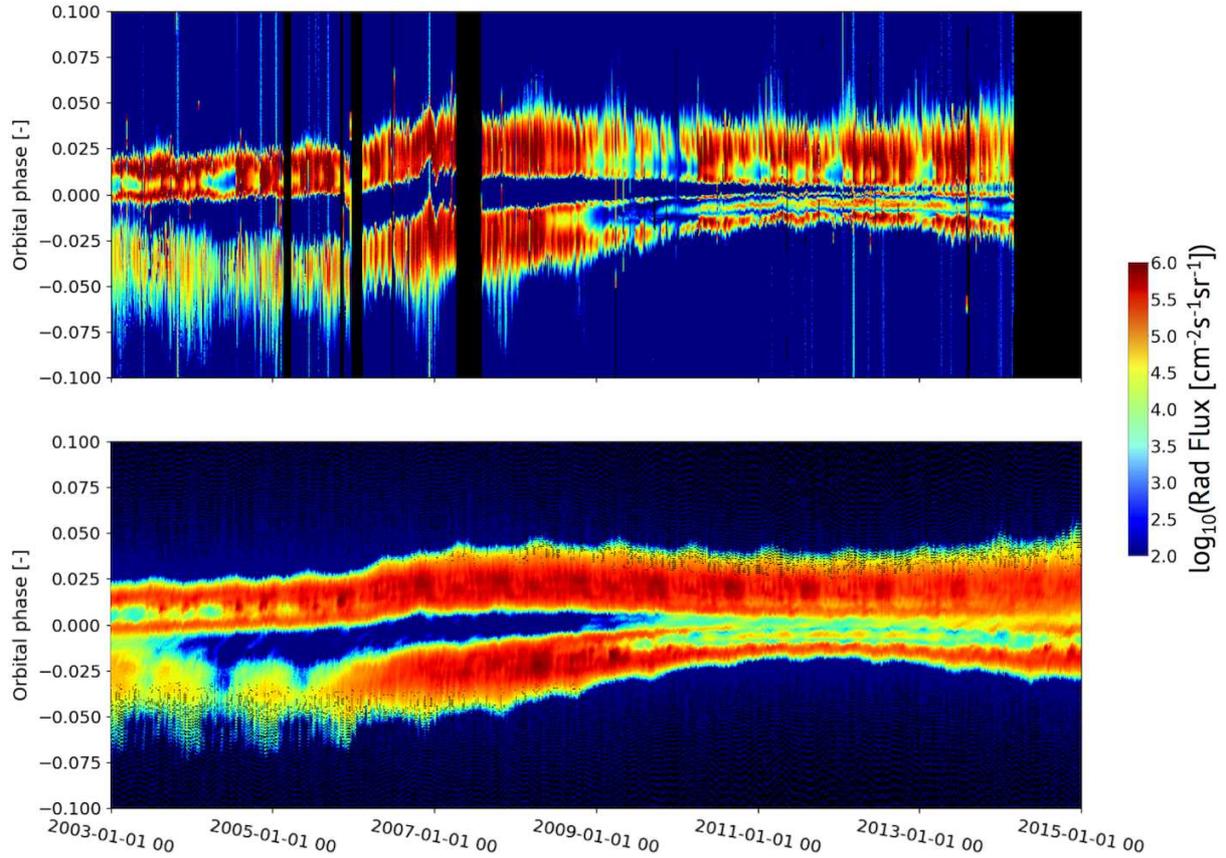}
    \caption{\integ mission radiation profiles between 2003 and 2015. Each vertical line corresponds to one revolution with the orbital phase 0 being perigee point. The top figure shows the real radiation data measurements. The second panel shows the radiation intensity extracted from the 5DRBM-e static model along \integ's trajectory. \label{fig:int2Dplots}}
\end{figure*}

\begin{figure*}[!t]
	\centering
    \includegraphics[width=0.87\textwidth]{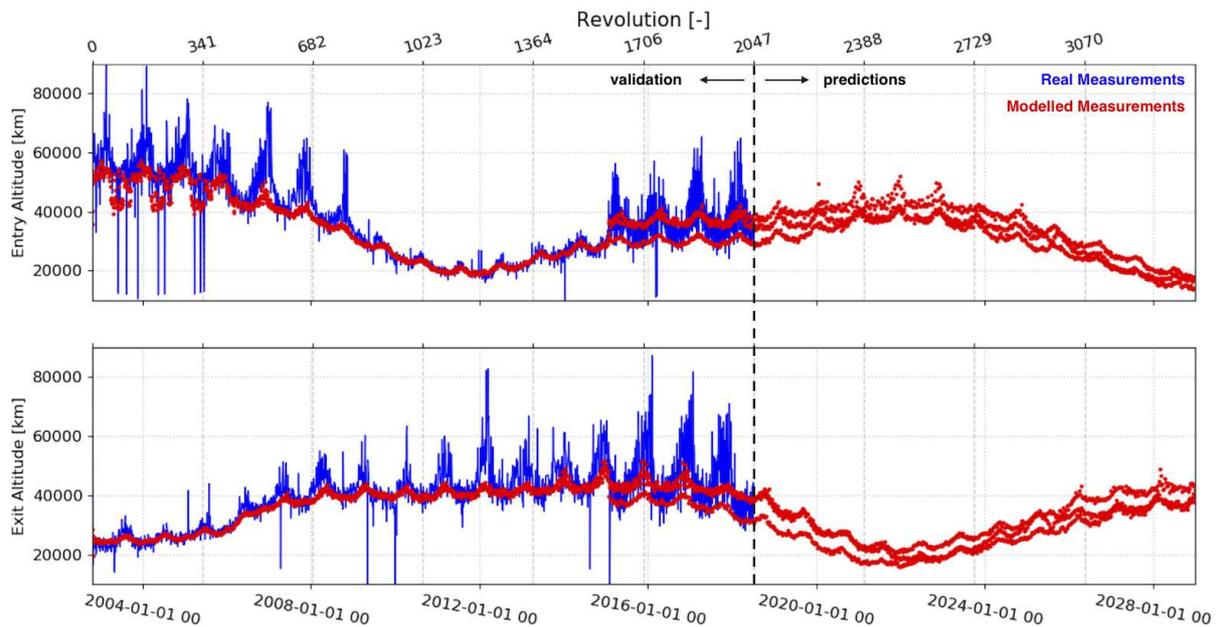}
    \caption{Belt entry and exit altitudes for each revolution for the whole \integ mission and for the expected future trajectory until the planned de-orbiting in 2029. The blue curves are the real measurements, and the red dots are the computed altitudes form the model based on a simple threshold on the electron counts.\label{fig:int_entextalt}}
\end{figure*}

\begin{figure*}[!t]
	\centering
    \includegraphics[width=0.87\textwidth]{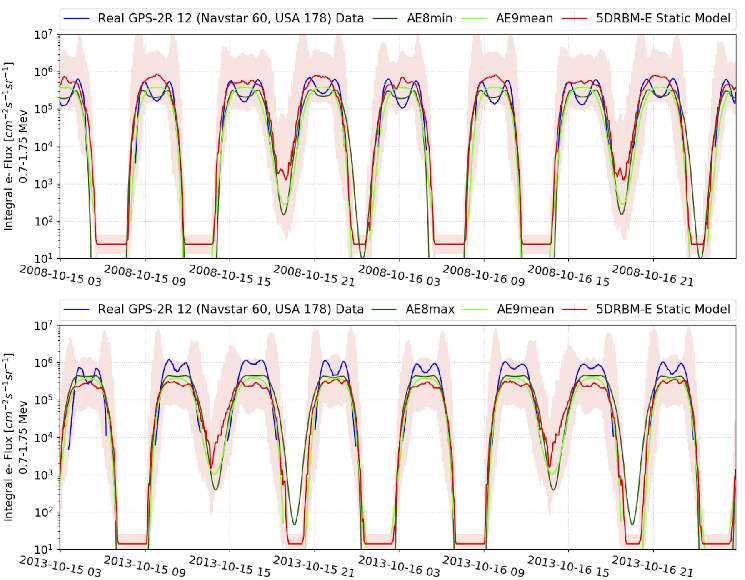}
    \caption{Radiation profiles for the GPS-2R 12 (Navstar 60, USA 178) satellite. The top figure shows the comparison of the real (blue), the AE8min (dark green), the AE9mean (light green) and the 5DRBM-e (red) radiation data during a low solar activity period. The bottom figure shows the radiation profile for the same satellite but for a high solar activity period and corresponding the AE8max model. The red-shaded region around the red model curve is the 95\% confidence interval. \label{fig:gpsae8ae9}}
\end{figure*}

The confidence on the model is estimated via the IQR in 3D cells of 1\rearth.
The IQR corresponds to a measure of dispersion of a data sample.
In each cell, the median radiation level of all data points is computed in order to split the values in the cell into two subsamples: one with the values below the median, and the other with values above the median.
For each subsample the medians are computed, and the IQR for the cell is given by the absolute difference between the medians of each subsample.
The confidence interval (uncertainty envelope) on the radiation profiles is computed from the IQRs across the modelled volume.
\cref{fig:IQRs} shows the $x-y$ and $x-z$ planes of the IQR map.
Looking at \cref{fig:step4} it can be seen that the largest dispersions are along the boundary of the VAB where the effects of the solar wind and solar activity are expected to be the strongest.
The IQRs could also be used to explore large time-dependent variations where values are the greatest, and checking time variations in each cell. This will be included in future work.

\subsection{The 5DRBM-e model}

The result is the 5DRBM-e model for trapped 0.7--1.75 MeV electrons.
It can be used with any orbit at any epoch to produce a predictive model of the radiation along the orbit as a function of the solar cycle.
The model is contained in a single FITS file that includes the static model, the sinusoidal fit parameters corresponding to the time-dependent solar cycle variations, and the IQR dispersion which can be used to calculate uncertainties and confidence intervals for any radiation profile.

\section{Validating the 5DRBM-e volume model}
\label{sec:model-validation}

\subsection{A self-consistency check}
\label{subsec:self-consistency-check}
A first step is a self-consistency check where the model predictions are compared to the actual data.
Naturally, the modelled radiation profile should be consistent with the data.
This comparison is shown in \cref{fig:int2Dplots} where it is seen that the model is not only consistent with the data, but that it accurately predicts radiation levels where there are data gaps (black regions in the top plot).

As mentioned in \cref{sec:radiation-monitors}, in order to avoid damaging their radiation-sensitive detectors, spacecraft like \xmm and \integ must turn off the scientific payload instruments before entering and turn them back on only after having exited the radiation belts.
Another check that can be done is to compare the model-predicted entry and exit altitudes to those measured.
This is shown in \cref{fig:int_entextalt} where measured altitudes are in blue, and predicted values are in red.
Long-term trends and seasonal variations are well captured, but not quite the peak amplitudes.
The main reason is the absence of time variations in the static model which smooths out the general shape of the belts.
The cross-calibration is also not perfect.
The split of the single red dotted curve into three in early 2015 is the consequence of the change in \integ's orbital period from 3 to 2.66 days following an orbital adjustment \citep{Dietze15}.
With an integer orbital period (3 days), \integ was scanning a similar part of the belt in each orbit. With the 2.66-day orbit, it needs 8 days (3 orbits) to come back to a similar region of the belt, and hence the three curves.

\subsection{Validation on GPS data}
\label{sec:GPS-check}
Actual validation must be done on data that have not been used to construct the model.
If, in addition, the result can be compared to the predictions of another model, then it is ideal.
The GNSS satellites orbits at approximately 20,000 km altitude are constantly inside the VAB, and thus subject to constant radiation.
This makes the accurate prediction of the radiation environment crucial to the success of any satellite on such orbits.
The AE8 and AP8 models are widely used for these particular regions of space as well as the newest AE9 model.
Hence, using the data from a GPS satellite and comparing the output of the 5DRBM-e model to that of the AE8 and AE9 models constitutes an ideal validation test.\footnote{The AE8 and AE9 models data are taken from the ESA Space Environment Information System (SPENVIS).}

\cref{fig:gpsae8ae9} shows a comparison between the static 5DRBM-e, the AE8min-AE8max and the AE9mean models for the GPS-2R 12 (Navstar 60, USA 178) satellite.
The measured radiation data and satellite positions for the dates shown in the figures are extracted from the ESA ODI server.
From these positions, a SPENVIS readable trajectory file is produced in order to compute the AE8 and AE9 modelled radiation data.
This GPS satellite is on a very low eccentricity orbit with an average altitude of 20,200 km and inclination of 55 degrees.
Both AE8min (during a solar minimum activity) and AE8max (during a solar maximum activity) as well as the AE9mean are compared to the static 5DRBM-e.
The GPS data, the AE8 and AE9 models are generally within the 95\% confidence interval (red-shaded region) of the 5DRBM-e model.

\section{Conclusion}

Modelling these still rather poorly understood radiation belts is important for space-borne activities (manned and unmanned) in the near-Earth environment.
The exploration of predictive methods used to ensure safe operations on \xmm and \integ, together with the benefit of creating global VAB models with great potentials, were the driving motivations for the work presented here.

The presented 5DRBM-e model has been built using \integ (>16 years) and \xmm (>18 years) radiation data, crossing the Van Allen Belts in each revolution.
This newly built data-driven model focuses on the electron belt from approximately 4,000 km up to its outer boundary at around 60,000 km.
The radiation flux given by the 5DRBM-e model corresponds to the integrated electron flux in the energy range from 0.7 to 1.75 MeV, the overlapping energy range for the radiation monitors on-board the two spacecraft.
Remarkably, not only have these two spacecraft gathered more than 16 years of contemporaneous radiation measurements, but their orbits scan different parts of the VAB and are thus complementary in increasing the coverage of the belts allowing the creation of a relevant global static model.

A reliable prediction of the radiation environment around the Earth requires an understanding of the radiation belts' dynamics.
Such knowledge is today surprisingly limited.
In this work, the time-dependent deviations from the static model are quantified using the interquartile range of radiation flux measurements in 3D cells over the entire volume defined in the model.
The IQR gives an excellent idea of the model's uncertainties on the radiation intensities, allowing, for example, the computation of confidence intervals for model-derived quantities such as the altitudes at entry and exit points.
This first static version of the 5DRBM-e with dynamic background radiation modelling, following the solar cycle, shows promising results with respect to the well-known AE8min and AE8max as well as AE9mean models.
In addition, a more accurate modelling of the VAB enhances the predictions of the belts entry and exit times which contributes to maximize the observation time and increase the safety during instrument operation. 
The simplicity of the 5DRBM-e model and its use of the Solar Magnetic reference frame make it easy to visualize in 3D space.
Moreover, the model has been structured in a step-by-step way with very few dependencies from one step to the another which makes it easy to update and modify.
If new data are available, a new updated and improved model can be built in a few hours only.
This versatility allows for the creation of models based on specific data sets with very few modifications to the procedure.

The intention is to extend this model to include more dynamic features associated with the 11-year solar cycle, as well as yearly variations that are observed in the data.
Short-term variations (days, weeks, months), mainly influenced by short-term solar events, will not be considered as many existing models already accomplish this, as outlined previously.
The electron energy range will also be increased by using more energy bins from \integ and \xmm and possibly from other missions.
Using more energy bins will require a good knowledge and extrapolation of the energy spectrum seen by both IREM and ERM instruments at each time-stamp, in order to compare the same integrated flux. The expected energy range should start at 0.7 MeV and go up to 2--3 MeV.
More importantly, a 5DRBM-p model will be constructed based on the proton radiation measurement data also available from the ODI server.
A first model will be built using only proton fluxes measured by \integ, because it has flown with a lower perigee altitude resulting in better coverage of the inner proton belt. The proton energy range will probably start at a few MeV up to several tens of MeV.

The end goal of this project is to provide robust, reliable, easy-to-use, data-driven, dynamic, 3D electron and proton radiation belt volume models that can be used in the space science and engineering community for designing, preparing, and running space missions.

%%%%%%%%%%%%%%%%%%%%%%%%%%%%%%%%%%%%%%%%%%%%%%%%%%%%%%%%%%%%%%%%%%%%%%%%%%%%%
%% Appendices
% \appendix

\section*{Acknowledgements}
\noindent
LM would like to thank the Swiss Space Centre (SSC), and the Swiss Space Office (SSO) for the financial support as a Swiss National Trainee at the European Space Astronomy Centre (ESAC/ESA).

%\clearpage

\end{document}